\documentclass[twocolumn,conference]{IEEEtran}

\usepackage{amsmath,amssymb,amsfonts}
\usepackage{graphicx}
\usepackage{textcomp}
\usepackage{xcolor}
\def\BibTeX{{\rm B\kern-.05em{\sc i\kern-.025em b}\kern-.08em
    T\kern-.1667em\lower.7ex\hbox{E}\kern-.125emX}}

\usepackage{hyperref}
\newcommand{\nix}[1]{}  

\begin{document}
\title{\LARGE{Detecting Glioma, Meningioma, and Pituitary Tumors, and Normal Brain Tissues based on Yolov11 and Yolov8 Deep Learning Models}}
\author{
Ahmed M. Taha$^a$,~  Salah A. Aly$^{b,c}$,~  Mohamed F. Darwish$^d$ \\
$^a$Dept. of CE, Faculty of Engineering,  Egypt University of Informatics, Cairo, Egypt    \\
$^b$Faculty of Computing and Data Science, Badya University, Giza, Egypt \\
$^c$CS\&Math Branch, Faculty of Science, Fayoum University, Fayoum, Egypt\\
$^d$Dept. of Pathology, Faculty of Medicine, Badya University, Giza, Egypt
    }
\maketitle

\begin{abstract}
Accurate and quick diagnosis of normal brain tissue Glioma, Meningioma, and Pituitary Tumors is crucial for optimal treatment planning and improved medical results. Magnetic Resonance Imaging (MRI) is widely used as a non-invasive diagnostic tool for detecting brain abnormalities, including tumors. However, manual interpretation of MRI scans is often time-consuming, prone to human error, and dependent on highly specialized expertise. This paper proposes an advanced AI-driven technique to detecting glioma, meningioma, and pituitary brain tumors using YoloV11 and YoloV8 deep learning models.

\smallskip
\noindent Methods: Using a transfer learning-based fine-tuning approach, we integrate cutting-edge deep learning techniques with medical imaging to classify brain tumors into four categories: No-Tumor, Glioma, Meningioma, and Pituitary Tumors. 

\smallskip
\noindent Results: The study utilizes the publicly accessible CE-MRI Figshare dataset and involves fine-tuning pre-trained models YoloV8 and YoloV11  of 99.49\% and  99.56\% accuraies;  and customized CNN accuracy of 96.98\%. The results validate the potential of CNNs in achieving high precision in brain tumor detection and classification, highlighting their transformative role in medical imaging and diagnostics. 
\end{abstract}


\section{Introduction}
Brain tumors represent one of the most critical challenges in modern medicine, owing to their complex detection and significant health implications. The advent of artificial intelligence (AI), particularly deep learning, has revolutionized the field of medical imaging by providing automated and highly accurate diagnostic systems. Convolutional Neural Networks (CNNs) have demonstrated exceptional performance in image classification tasks, enabling advancements in tumor detection and classification. Despite the success of pre-trained models such as ResNet, VGG, and MobileNet, questions remain about their optimal use and whether custom-built deep neural networks (DNNs) can surpass them in specific tasks like brain tumor classification.

This study is dedicated to addressing critical questions in the field of brain tumor detection by conducting a comprehensive evaluation of various pre-trained models, including   YOLO8, and YOLOv11, alongside custom-designed deep neural networks. The evaluation is carried out on multiple brain tumor datasets, with the overarching aim of identifying the most effective model for accurate identification and classification of brain tumors.

A primary objective of this research is to assess the diagnostic performance of the Yolov8 and Yolov8  models in real-world scenarios and establish benchmarks for their accuracy, reliability, and computational efficiency. Additionally, this study seeks to explore and implement optimization techniques that can further enhance the diagnostic capabilities of the models under investigation.

\begin{figure}[h]
\centerline{\includegraphics[width=9cm, height=6cm]{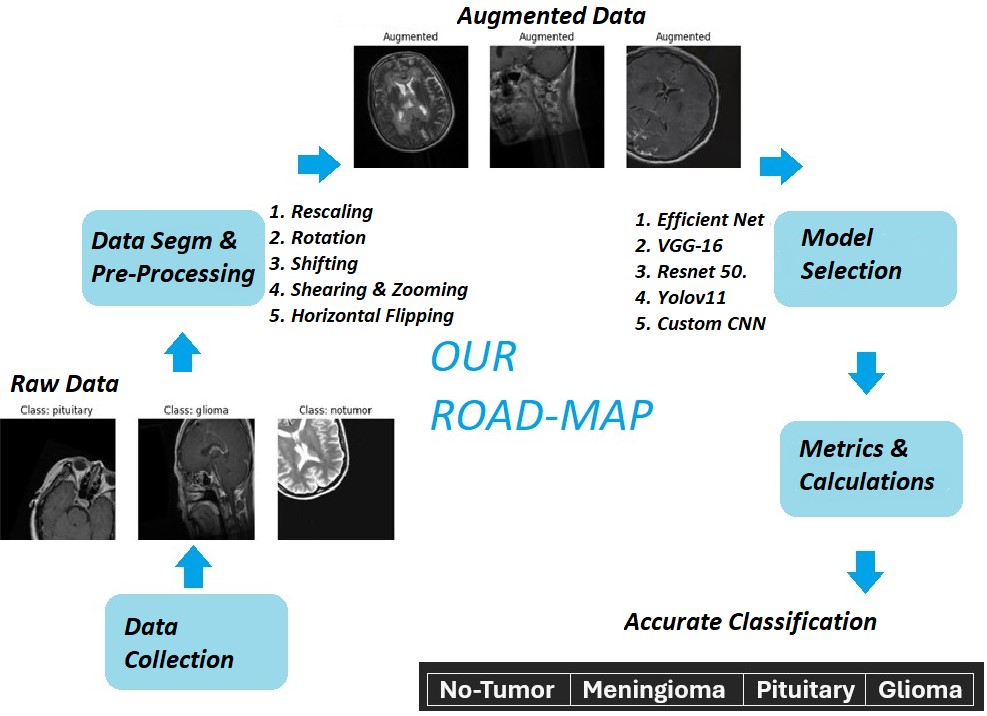}}
\caption{The framework of our study}\label{fig}
\end{figure}

A detailed analysis of comparison with different models will enable the development of strategies to overcome challenges and improve overall performance. Through this process, the work aspires to contribute to the development of robust, reliable, and efficient diagnostic tools for brain tumor detection, ultimately aiding in early and accurate medical diagnoses and improving patient outcomes. The main contributions of this work are stated as follows:
\begin{enumerate}
  \item Proving that Optimized CNN Design can Surpass Pre-Trained Models for Brain Tumor Classification
  \item YoloV11 and YoloV8 achieves Exceptional Accuracy in Brain Tumor Classification

\end{enumerate}

The remainder of this paper is organized as follows.    Section~\ref{sec:datasets}  provides the datasets description. Section~\ref{sec:models}  presents into the specifics of the models employed in this research.  Sections~\ref{sec:metrics}, \ref{sec:resultsYolo}, \ref{sec:resultsCNN}  present the performance metric and results of the pre-trained models. Section~\ref{sec:related} presents an in-depth review of the most recent and well-established studies in the field of brain tumor classification using machine learning and deep learning techniques.  Section \ref{sec:conclusion} summarizes and concludes this paper.


\section{Datasets and Data Collection}\label{sec:datasets}
For this study, we curated a comprehensive dataset by filtering and combining three of the most recently published datasets relevant to brain tumor detection and classification. This integrated dataset was specifically designed to address our research objectives and provide a robust foundation for model training and evaluation. The curated dataset is organized into four distinct categories: No-Tumor, Glioma, Meningioma, and Pituitary. these three data sets are as follows, to see samples of the data Please follow up the link in the reference:

        \begin{table}[htbp]
        \caption{1st Dataset with $7023$ images}
        \centering
        \begin{tabular}{|p{1.5cm}|p{1.3cm}|p{1cm}|p{1cm}|p{1cm}|}
        \hline
        \textbf{Class} & \textbf{No. Images} & \textbf{Percent.} & \textbf{Training 80\%} & \textbf{Testing 20\%}\\
        \hline\hline
        No-Tumor & 2000 & 28.4\% & 1595 & 405\\
        Glioma   & 1621 & 23\% & 1321 & 300\\
        Meningioma      & 1645 & 23.4\%  & 1339 & 306\\
        Pituitary      & 1757 & 25\%& 1457 & 300\\
              \hline
        \end{tabular}
        \label{tab:Analysis2}
        \end{table}
The first dataset is published oct 2024\cite{ds1}, Among the selected datasets, one prominent source is the Brain Tumor Classification Dataset on Kaggle, it has 7024 images divided into the aforementioned four categories. This dataset comprises a substantial collection of Magnetic Resonance Imaging (MRI) scans of the brain. Notably, the MRI images in this dataset are presented in their raw form, without any pre-processing techniques applied, thus preserving the original data distribution and characteristics.

The second one is published Nov 2024\cite{ds2}, To enhance the robustness and diversity of our dataset, we incorporated an additional publicly available dataset containing a significantly larger collection of images. This supplementary dataset comprises a total of 30,700 MRI images, systematically categorized into the aforementioned 4 categories.
        \begin{table}[htbp]
        \caption{2nd Dataset with  $30.100$ images}
        \centering
        \begin{tabular}{|p{1.5cm}|p{1.3cm}|p{1cm}|p{1cm}|p{1cm}|}
        \hline
        \textbf{Class} & \textbf{No. Images} & \textbf{Percent.} & \textbf{Training 100\%}\\
        \hline\hline
        No-Tumor & 5588 & 18.5\% & 5588\\
        Glioma   & 7928 & 26.3\% & 7928\\
        Meningioma      & 8900 & 29.5\% & 8900\\
        Pituitary      & 7665 & 25.4\% & 7665\\
        \hline
        \end{tabular}
        \label{tab:Analysis2}
        \end{table}

The final dataset is published Dec 2024 \cite{ds3}, It was carefully selected to prioritize image quality and its impact on model performance. Although this dataset represents the smallest portion of our overall collection, it is of critical importance due to the high-resolution Magnetic Resonance Imaging (MRI) scans it contains. These high-quality images played a pivotal role in enhancing the accuracy and reliability of our deep learning models. This dataset comprises a total of 2,107 images, systematically divided into two primary subsets: Training and Testing. Each of these subsets is further categorized into the same four classes as previously described: No-Tumor, Glioma, Meningioma, and Pituitary.
        \begin{table}[htbp]
        \caption{3rd Dataset with  $2107$ images}
        \centering
        \begin{tabular}{|p{1.5cm}|p{1.3cm}|p{1cm}|p{1cm}|p{1cm}|}
        \hline
        \textbf{Class} & \textbf{No. Images} & \textbf{Percent.} & \textbf{Training 80\%} & \textbf{Testing 20\%}\\
        \hline\hline
        No-Tumor & 558 & 26.4\% & 437 & 121\\
        Glioma   & 487 & 23.1\% & 397 & 90\\
        Meningioma      & 493 & 23.4\% & 401 & 92\\
        Pituitary      & 563 & 26.7\% & 473 & 90\\
        \hline
        \end{tabular}
        \label{tab:Analysis2}
        \end{table}

The integration of this dataset was motivated by the need to expand the volume and variability of the training and testing samples, thereby improving the generalizability of our deep learning models. By including a larger number of images across all categories, this augmentation step ensures that the models are exposed to a wide range of anatomical variations, imaging conditions, and potential noise, which is crucial for developing a reliable and accurate classification system.

Table~\ref{tab:Analysis2} shows the class breakdown, including the training, validation, and test splits.

\section{Methodology}\label{sec:models}
This section begins by providing a detailed overview of the historical development, theoretical foundations, and structural design of the pre-trained models utilized in the study, such as YOLOv8, and YOLOv11.

\subsection{Data Pre-Processing}
To enhance the quality and ensure precise positional alignment of the dataset images, we employed a comprehensive series of preprocessing steps designed to address common challenges in medical imaging data. These steps were specifically tailored to mitigate the effects of noise, improve image clarity, and emphasize the critical features necessary for accurate brain tumor classification. By refining the visual quality and structural integrity of the input data, these enhancements played a pivotal role in boosting the performance of our deep learning models. Such preprocessing not only facilitated better feature extraction during the training phase but also contributed to improved model efficiency by reducing unnecessary variability in the data, these preprocessing steps are as follows:

\noindent\textbf{Re-scaling:} MRI images often contain pixel intensity values that vary widely across different scans [0 - 255]. Scaling ensures that these intensity values are normalized to a consistent range [0 - 1], which helps prevent numerical instability during model training and ensures uniformity across input data, The re-scaling process is also very important at Facilitation of Model Convergence.
   
\noindent\textbf{Rotation:} The images are rotated randomly by an angle chosen between -30 and +30 degrees. For example, an image can be tilted slightly clockwise or counterclockwise, Adding these rotational variations makes the model robust to different orientations of the same object. This ensures the model doesn't overly rely on the object's alignment during training.

 \noindent\textbf{Shifting (Horizontal \& Vertical):} The image is randomly shifted horizontally (left or right) or vertically (up or down) by a maximum of 30\% of the image dimensions, It simulates real-world scenarios where objects may not always be perfectly centered in the frame, helping the model generalize better.

\noindent \textbf{Shearing \& Zooming:} Shearing involves distorting the image such that it appears stretched in a particular direction, as if it’s being viewed from a tilted angle, This simulates perspective distortions that can occur in real-world images. Whereas Zooming The image is randomly zoomed in or out by up to ±20\% of its size, it ensures the model learns to identify features at varying scales, enhancing its ability to recognize objects at different distances.

\noindent\textbf{Horizontal Flipping:} : Randomly flipping images horizontally, The image is flipped along its vertical axis, effectively mirroring it. For example, an image of a cat facing left becomes an image of the cat facing right, It increases dataset diversity, especially in cases where the orientation of the object (left or right) is not significant, such as in recognizing animals or faces.

\subsection{Image Classification Models}\label{models}
In this section, we delve into a comprehensive analysis of four widely recognized pre-trained models YOLOv8, and YOLOv11 alongside a customized Convolutional Neural Network (CNN) architecture. These models were carefully selected due to their established performance and versatility in addressing various computer vision tasks, including medical image analysis. Our goal is to evaluate and compare their effectiveness in classifying brain tumor types using the refined datasets derived from the previously highlighted three data sources.
Each of these pre-trained models brings unique architectural strengths and design philosophies, making them suitable for different aspects of medical imaging and diagnostic applications.

\subsection{YOLO}
 Based on these references \cite{yoloRef1}, \cite{redmon2016lookonceunifiedrealtime}, and \cite{yoloRef2} Joseph Redmon, Santosh Divvala, Ross Girshick, Ali Farhadi published a paper named “You Only Look Once: Unified, Real-Time Object Detection” at CVPR, introducing a revolutionary model named YOLO. In our study, we are going to use both YOLO-8 and YOLO-11 which is the latest version of YOLO family, It introduces a more efficient architecture with C3K2 blocks, SPFF (Spatial Pyramid Pooling Fast), and advanced attention mechanisms like C2PSA. YOLOv11 is designed to enhance small object detection and improve accuracy while maintaining the real-time inference speed that YOLO is known form, its backbone is known as Conv Block which process the given c,h,w passing through a 2D convolutional layer following with a 2D Batch Normalization layer at last with a SiLU Activation Function rather than Sigmoid function.
            The tuning processes for YOLO v8 and YOLO v11 involved several phases that were similar to those applied to the previous models. The specific characteristics and properties of their training processes are detailed in the table below.
            \begin{table}[htbp]
            \caption{YOLOv8 \& YOLOv11 training characteristics}
            \centering
            \begin{tabular}{|p{1.8cm}|p{1.4cm}|p{1.4cm}|}
            \hline
            Epochs	&20\\
            \hline
            Image Size	& 224x224\\
            \hline
            Batch Size	& 32\\
            \hline
            Optimizer   & Adam\\
            \hline
            Learning Rate   & 0.001\\
            \hline
            Augmentation    & Enabled\\
            \hline
            Device      & GPU (device0)\\
            \hline
            \end{tabular}
            \label{tab:yoloresults}
            \end{table}

\section{Performance Metrics}\label{sec:metrics}
Numerous evaluation methods and metrics have been developed to assess different types of tasks in deep learning. In this study, we have used several metrics to illustrate and comprehend the efficiency of the models used. These calculations provided us with a clear explanation for the results obtained from each model during the experimentation phase, which assisted and guided us when tweaking the methodologies for optimal results. For this purpose, accuracy, F1-score, precision, recall, specificity, and confusion matrix were computed.

Accuracy is an overall indicator of how well the model performs, considering the number of correctly identified samples out of all the given samples. It is represented by the sum of true positives and true negatives divided by the total number of examples, which includes True Positive (TP), True Negative (TN), False Positive (FP), and False Negative (FN), as expressed in Equation~\ref{eq:acc}:
\begin{equation}
 Accuracy = \frac{TP+TN}{TP+TN+FP+FN}
 \label{eq:acc}
\end{equation}

Presented in Equation~\ref{eq:prec}, the sample precision which is identified by the ratio of correctly classified instances to the total number of classified instances.
\begin{equation}
 Precision = \frac{TP}{TP+FP}
 \label{eq:prec}
\end{equation}

Recall, or Sensitivity, is calculated as the ratio of correctly
identified instances to the total number of instances, as described
in Equation~\ref{eq:recall}:
\begin{equation}
 Recall = \frac{TP}{TP+FN}
 \label{eq:recall}
\end{equation}

Another significant metric that contributed to our results is
F1-score. It is obtained by calculating the harmonic mean of
precision and recall, as illustrated in Equation~\ref{eq:f1}:
\begin{equation}
 F1-score = 2 \times \frac{Percision \times Recall}{Percision + Recall}
 \label{eq:f1}
\end{equation}

\section{YOLOv8 \& YOLOv11 Experimental Results}\label{sec:resultsYolo}
After thoroughly describing the architecture of YOLOv8 and YOLOv11, which are optimized for real-time object detection and classification tasks, we highlighted the fine-tuning strategies and customizations applied to enhance their performance for brain tumor classification, see~\cite{Awad2024} for utilizing YOLOv8 and YOLOv11 in  similar scenarios. We also outlined the specific training parameters and configurations tailored for these models. Following this, the performance of YOLOv8 and YOLOv11 was evaluated using the four critical metrics: accuracy, precision, recall, and F1-score. These metrics demonstrate the models’ exceptional ability to process MRI images effectively and classify brain tumors with remarkable speed and accuracy, underscoring YOLO's suitability for medical imaging tasks. in the following 2 tables, table \ref{tab:yolo8 results} displays the results of YOLOv8 whereas table \ref{tab:yolo11 results} displays the results of YOLOv11.
\begin{table}[htbp]
\caption{YOLOv8 Results}
\centering
\begin{tabular}{|p{1.8cm}|p{1.4cm}|p{1.4cm}|}
\hline
Metric	& Training	& Validation \\
\hline
Accuracy	& 99.69\%	& 99.31\%\\
\hline
Precision	& 99.10\%	& 98.69\%\\
\hline
Recall	& 99.42\%& 	99.13\%\\
\hline
F1-Score& 	98.88\%	& 98.50\%\\
\hline
\end{tabular}
\label{tab:yolo8 results}
\end{table}

\begin{figure}[!h]
        \centerline{\includegraphics[width=8.5cm, height=4cm] {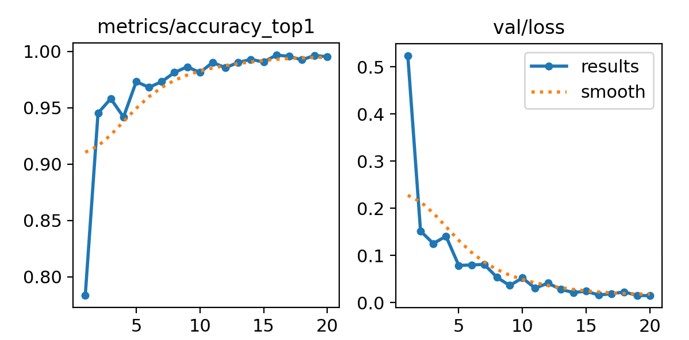}}
        \caption{YOLOv8 Accuracy \& Loss}
        \label{fig:yolov8graph}
        \end{figure}

\begin{figure}[!h]
        \centerline{\includegraphics[width=8.5cm, height=4cm] {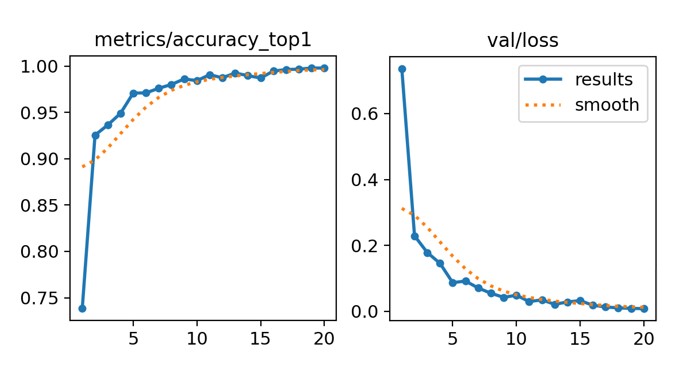}}
        \caption{YOLOv11 Accuracy \& Loss}
        \label{fig:yolov11graph}
        \end{figure}

\begin{table}[htbp]
\caption{YOLOv11 Results}
\centering
\begin{tabular}{|p{1.8cm}|p{1.4cm}|p{1.4cm}|}
\hline
Metric	& Training	& Validation \\
\hline
Accuracy	& 99.89\%	& 99.50\%\\
\hline
Precision	& 99.52\%	& 99.11\%\\
\hline
Recall	& 99.62\%& 	99.24\%\\
\hline
F1-Score& 	99.36\%	& 98.15\%\\
\hline
\end{tabular}
\label{tab:yolo11 results}
\end{table}
Figure \ref{fig:yolov8graph} and Figure \ref{fig:yolov11graph} shows the accuracy and loss for validation data for both YOLOv8 and YOLOv11 in sequence.
        
\begin{figure}[!h]
        \centerline{\includegraphics[width=8.5cm, height=5cm] {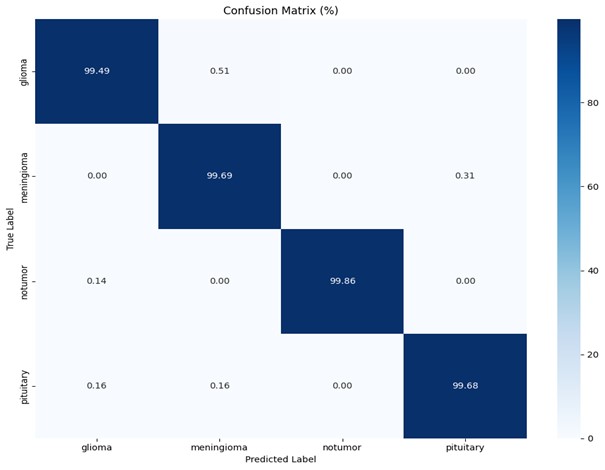}}
        \caption{YOLOv8 Confusion Matrix}
        \label{fig:yolov8conf}
        \end{figure}

\begin{figure}[!h]
        \centerline{\includegraphics[width=8.5cm, height=5cm] {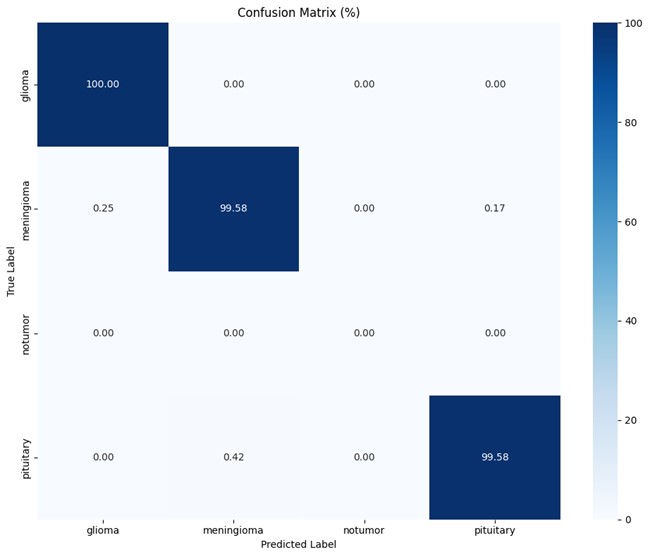}}
        \caption{YOLOv11 Confusion Matrix}
        \label{fig:yolov11conf}
        \end{figure}

Both figure \ref{fig:yolov8conf} and figure \ref{fig:yolov11conf} display the confusion matrixes after fine tuning the parameters of YOLOv8 and YOLOv11 in sequence.
        
\section{Customized CNN  Experimental Results}\label{sec:resultsCNN}
After detailing the design of our customized CNN model, including its architecture, the number of convolutional layers, activation functions, dropout rates, optimizer configurations, and loss function selection, we evaluated its performance on the brain tumor classification task. The model's training was optimized for our dataset, with all parameters fine-tuned to achieve high precision and reliability. The results were analyzed using four critical metrics: accuracy, precision, recall, and F1-score, which demonstrated the model's ability to achieve superior results compared to some pre-trained models, showcasing the potential of a carefully tailored CNN architecture for specialized medical imaging tasks.

\textbf{\textit{Notably:}} the results obtained from the CNN model represent one of the most significant deliverables of this research. These findings underscore the potential of a well-designed and carefully customized CNN architecture to surpass the performance of many pre-trained models. By tailoring the CNN to the specific requirements of the task and optimizing its structure, the model demonstrates remarkable efficiency and accuracy. This advantage arises from the fact that the CNN is trained on its raw, unaltered data, free from any extraneous noise or biases typically present in pre-trained models. This highlights the flexibility and adaptability of CNN architectures when meticulously designed for domain-specific applications, such as brain tumor classification.
\begin{table}[htbp]
\caption{CNN Results}
\centering
\begin{tabular}{|p{1.8cm}|p{1.4cm}|p{1.4cm}|}
\hline
Metric	& Training	& Validation \\
\hline
Accuracy	& 98.50\%	& 97.01\%\\
\hline
Precision	& 98.32\%	& 97.41\%\\
\hline
Recall	& 97.63\%& 	96.81\%\\
\hline
F1-Score& 	98.12\%	& 97.16\%\\
\hline
\end{tabular}
\label{tab:cnn results}
\end{table}

\begin{figure}[!h]
        \centerline{\includegraphics[width=8.5cm, height=4cm] {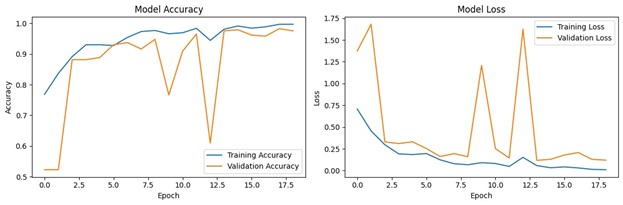}}
        \caption{CNN Accuracy \& Loss for Training and Validation}
        \label{fig:cnngraph}
        \end{figure}

Figure \ref{fig:cnngraph} shows the accuracy and loss for training and validation data for the CNN model.
        
\begin{figure}[!h]
        \centerline{\includegraphics[width=8.5cm, height=5cm] {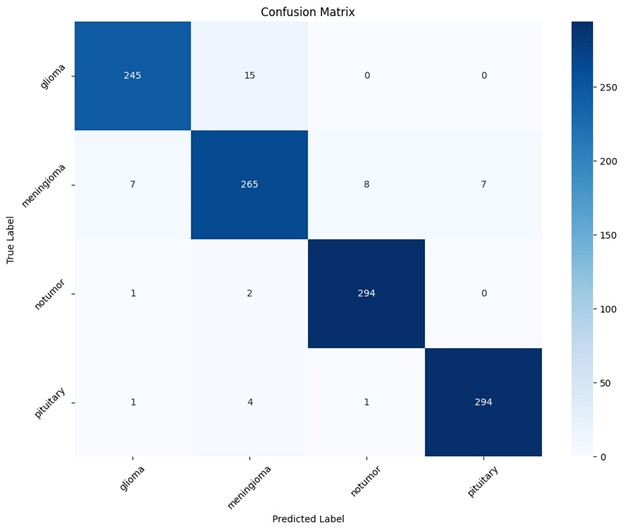}}
        \caption{CNN Confusion Matrix}
        \label{fig:cnnconf}
        \end{figure}

Figure \ref{fig:cnnconf} displays the confusion matrixes after fine tuning the parameters of CNN.

\section{Related Work and Comparison Study}\label{sec:related}

 P.S. Smitha and G. Balaarunesh~\cite{SMITHA2024101295} proposed a comprehensive deep learning approach for early brain tumor classification using medical imaging data. A diverse dataset encompassing various tumor types, stages, and healthy brain images is utilized. Preprocessing techniques like augmentation and normalization enhance data robustness. A convolutional neural network (CNN) architecture serves as the primary model, leveraging transfer learning from pre-trained models to extract relevant features even with limited data. The training process optimizes hyperparameters to prevent overfitting, and performance is evaluated using metrics like accuracy, precision, recall, F1-score, confusion matrices, and ROC curves on a separate test set.

 Naeem Ullah and Ali Javed~\cite{10.1371/journal.pone.0291200} They proposed a novel unified end-to-end deep learning model named TumorDetNet for brain tumor detection and classification. Their TumorDetNet framework employs 48 convolution layers with leaky ReLU (LReLU) and ReLU activation functions to compute the most distinctive deep feature maps. Moreover, average pooling and a dropout layer are also used to learn distinctive patterns and reduce overfitting. Finally, one fully connected and a softmax layer are employed to detect and classify the brain tumor into multiple types.

 Sonia Ben Brahim and Samia Dardouri~\cite{https://doi.org/10.1155/2024/7634426} Their study involved the application of a deep convolutional neural network (DCNN) to diagnose brain tumors from MR images. The application of these algorithms offers several benefits, including rapid brain tumor prediction, reduced errors, and enhanced precision. The proposed model is built upon the state-of-the-art CNN architecture VGG16, employing a data augmentation approach

 Md Zehan Alam and Tonmoy Roy~\cite{alam2024enhancingtransferlearningmedical} Explored and enhances the application of Transfer Learning (TL) for multilabel image classification in medical imaging, focusing on brain tumor class and diabetic retinopathy stage detection. The effectiveness of TL—using pre-trained models on the ImageNet dataset—varies due to domain-specific challenges. They evaluate five pretrained models—MobileNet, Xception, InceptionV3, ResNet50, and DenseNet201—on two datasets: Brain Tumor MRI and APTOS 2019. Their results show that TL models excel in brain tumor classification, achieving near-optimal metrics.

   Abdullah al Nomaan Nafi and Md. Alamgir Hossain~\cite{nafi2024diffusionbasedapproachesmedicalimage} Investigated the effectiveness of diffusion models for generating synthetic medical images to train CNNs in three domains: Brain Tumor MRI, Acute Lymphoblastic Leukemia (ALL), and SARS-CoV-2 CT scans. A diffusion model was trained to generate synthetic datasets for each domain. Pre-trained CNN architectures were then trained on these synthetic datasets and evaluated on unseen real data.

   Bijay Adhikari and Pratibha Kulung~\cite{adhikari2024parameterefficientfinetuningimprovedconvolutional} They proposed Convolutional adapter-inspired Parameter-efficient Fine-tuning (PEFT) of MedNeXt architecture. To validate their idea, They show their method performs
    comparable to full fine-tuning with the added benefit of reduced training     compute using BraTS-2021 as pre-training dataset and BraTS-Africa
    as the fine-tuning dataset. BraTS-Africa consists of a small dataset (60    train / 35 validation) from the Sub-Saharan African population with
    marked shift in the MRI quality compared to BraTS-2021 (1251 train    samples). They first show that models trained on BraTS-2021 dataset do
    not generalize well to BraTS-Africa as shown by 20\% reduction in mean    dice on BraTS-Africa validation samples. Then, They show that PEFT
    can leverage both the BraTS-2021 and BraTS-Africa dataset to obtain    mean dice of 0.8 compared to 0.72 when trained only on BraTS-Africa.
    Finally, They show that PEFT (0.80 mean dice) results in comparable performance    to full fine-tuning (0.77 mean dice) which may show PEFT to
    be better on average but the boxplots show that full finetuning results     is much lesser variance in performance.
    
   Abhijeet Parida and Daniel Capellán-Martín~\cite{parida2024adultgliomasegmentationsubsaharan} They leverage pre-trained deep learning models, nnU-Net and MedNeXt, and apply a stratified fine-tuning strategy using the BraTS2023-Adult-Glioma and BraTS-Africa datasets. Their method exploits radiomic analysis to create stratified training folds, model training on a large brain tumor dataset, and transfer learning to the Sub-Saharan context. A weighted model ensembling strategy and adaptive post-processing are employed to enhance segmentation accuracy. The evaluation of Their proposed method on unseen validation cases on the BraTS-Africa 2024 task resulted in lesion-wise mean Dice scores of 0.870, 0.865, and 0.926, for enhancing tumor, tumor core, and whole tumor regions and was ranked first for the challenge. Their approach highlights the ability of integrated machine-learning techniques to bridge the gap between the medical imaging capabilities of resource-limited countries and established developed regions. 

The main contributions of this work are stated as follows:\\
\textbf{Fusion of Optimized Pre-Trained Models:} The research systematically evaluates and fine-tunes multiple pre-trained models (e.g., YOLOv11 and Yolov8) for brain tumor classification, providing a comparative analysis of their performance. This fusion approach sets a benchmark for integrating and optimizing pre-trained models in medical imaging.
    
\textbf{Integrating Recent Brain Tumor Datasets:} By incorporating diverse, high-quality MRI datasets, the study enhances the model's generalization capabilities, reducing overfitting and improving performance across varied real-world scenarios.

\begin{table}[ht]
    \centering
    \caption{Comparison of Different Approaches for Detecting brain tumor detections}
   \begin{tabular}{ |p{1.5cm}|p{1.8cm}|p{1.9cm}|p{1.6cm}|}
        \hline \hline
               \textbf{Research Name} & \textbf{Dataset}& \textbf{Methodology} & \textbf{Accuracy}\\
        & &  & \\
        \hline 
          Classification of Brain Tumor  & BraTS &- Mask R-CNN & 92.3 \\
        
          & &- SVM &  85.4\\
          & &- Random Forest & 82.7\\
        \hline
       Diffusion-Based Approaches  & Brain Tumor MRI,  SARS-CoV-2 CT-Scans&  - ResNet-50& 86.94\\
         & &- VGG-16  &   83.01\\
         & &- MobileNetV2 &   85.21\\
         & &- GoogleNet &   89.40\\
         & &- AlexNet &   82.22\\
        \hline
       Our Research   & Hybrid dataset extracted from 3 different datasets &- YOLOv8  &99.49\\
          & &- YOLOv11 & 99.56\\
          & &- CNN &  97.01\\
          \hline
    \end{tabular}
    \label{tab:ALL_detection_comparison}
\end{table}

\section{Conclusion}\label{sec:conclusion}
We highlight that YOLO remains one of the most robust and reliable pre-trained architectures for detecting brain tumors, showcasing its superior performance in this domain.    We demonstrated that a well-designed and carefully customized CNN architecture can outperform many pre-trained models. This is attributed to its ability to train directly on raw, clean, and domain-specific data without incorporating the unnecessary noise or unrelated features often inherent in pre-trained models.     Our results confirmed that YOLOv8 and YOLOv11 stand out as the most effective pre-trained models for MRI classification tasks, particularly in brain tumor detection. These models showcased exceptional accuracy and robustness, setting a benchmark for medical imaging applications.

\bibliographystyle{IEEEtran}
\bibliography{braincancer_refer}

\begin{thebibliography}{10}
\providecommand{\url}[1]{#1}
\csname url@samestyle\endcsname
\providecommand{\newblock}{\relax}
\providecommand{\bibinfo}[2]{#2}
\providecommand{\BIBentrySTDinterwordspacing}{\spaceskip=0pt\relax}
\providecommand{\BIBentryALTinterwordstretchfactor}{4}
\providecommand{\BIBentryALTinterwordspacing}{\spaceskip=\fontdimen2\font plus
\BIBentryALTinterwordstretchfactor\fontdimen3\font minus
  \fontdimen4\font\relax}
\providecommand{\BIBforeignlanguage}[2]{{%
\expandafter\ifx\csname l@#1\endcsname\relax
\typeout{** WARNING: IEEEtran.bst: No hyphenation pattern has been}%
\typeout{** loaded for the language `#1'. Using the pattern for}%
\typeout{** the default language instead.}%
\else
\language=\csname l@#1\endcsname
\fi
#2}}
\providecommand{\BIBdecl}{\relax}
\BIBdecl

\bibitem{ds1}
\BIBentryALTinterwordspacing
M.~S. Gado, ``Brain tumor classification ds1,'' 2024. [Online]. Available:
  \url{https://www.kaggle.com/datasets/mahmoudsalamagado/brain-tumor}
\BIBentrySTDinterwordspacing

\bibitem{ds2}
\BIBentryALTinterwordspacing
M.~Hamada, ``Brain tumor classification ds2,'' 2024. [Online]. Available:
  \url{https://www.kaggle.com/datasets/mohamada2274/brain-tumor-mri}
\BIBentrySTDinterwordspacing

\bibitem{ds3}
\BIBentryALTinterwordspacing
APPASAMI.G, ``Brain tumor classification ds3 for accuracy,'' 2024. [Online].
  Available: \url{https://www.kaggle.com/datasets/appasamig/mri-brain-tumor-lr}
\BIBentrySTDinterwordspacing

\bibitem{yoloRef1}
\BIBentryALTinterwordspacing
R.~Khanam* and M.~Hussain, ``Yolov11: An overview of the key architectural
  enhancements,'' 2024. [Online]. Available:
  \url{https://arxiv.org/html/2410.17725v1?utm_source=chatgpt.com}
\BIBentrySTDinterwordspacing

\bibitem{redmon2016lookonceunifiedrealtime}
\BIBentryALTinterwordspacing
J.~Redmon, S.~Divvala, R.~Girshick, and A.~Farhadi, ``You only look once:
  Unified, real-time object detection,'' 2016. [Online]. Available:
  \url{https://arxiv.org/abs/1506.02640}
\BIBentrySTDinterwordspacing

\bibitem{yoloRef2}
\BIBentryALTinterwordspacing
A.~T. Khan, ``A unified framework for leaf extraction and analysis in
  multi-crop phenotyping using yolov11,'' 2025. [Online]. Available:
  \url{https://www.researchgate.net/publication/388129085_LEAF-Net_A_Unified_Framework_for_Leaf_Extraction_and_Analysis_in_Multi-Crop_Phenotyping_Using_YOLOv11}
\BIBentrySTDinterwordspacing

\bibitem{Awad2024}
A.~Awad and S.~A. Aly, ``Early diagnosis of acute lymphoblastic leukemia using
  yolov8 and yolov11 deep learning models,'' in \emph{IEEE JAC-ECC,
  International Japan-Africa Conference on Electronics communications and
  Computations, Alex, 16-18 December, 2024. arXiv preprint arXiv:2410.10701},
  2024.

\bibitem{SMITHA2024101295}
\BIBentryALTinterwordspacing
P.~Smitha, G.~Balaarunesh, C.~{Sruthi Nath}, and A.~{Sabatini S},
  ``Classification of brain tumor using deep learning at early stage,''
  \emph{Measurement: Sensors}, vol.~35, p. 101295, 2024. [Online]. Available:
  \url{https://www.sciencedirect.com/science/article/pii/S266591742400271X}
\BIBentrySTDinterwordspacing

\bibitem{10.1371/journal.pone.0291200}
\BIBentryALTinterwordspacing
N.~Ullah, A.~Javed, A.~Alhazmi, S.~M. Hasnain, A.~Tahir, and R.~Ashraf,
  ``Tumordetnet: A unified deep learning model for brain tumor detection and
  classification,'' \emph{PLOS ONE}, vol.~18, no.~9, pp. 1--24, 09 2023.
  [Online]. Available: \url{https://doi.org/10.1371/journal.pone.0291200}
\BIBentrySTDinterwordspacing

\bibitem{https://doi.org/10.1155/2024/7634426}
\BIBentryALTinterwordspacing
S.~Ben~Brahim, S.~Dardouri, and R.~Bouallegue, ``Brain tumor detection using a
  deep cnn model,'' \emph{Applied Computational Intelligence and Soft
  Computing}, vol. 2024, no.~1, p. 7634426, 2024. [Online]. Available:
  \url{https://onlinelibrary.wiley.com/doi/abs/10.1155/2024/7634426}
\BIBentrySTDinterwordspacing

\bibitem{alam2024enhancingtransferlearningmedical}
\BIBentryALTinterwordspacing
M.~Z. Alam, T.~Roy, H.~M.~N. Kawsar, and I.~Rimi, ``Enhancing transfer learning
  for medical image classification with smote: A comparative study,'' 2024.
  [Online]. Available: \url{https://arxiv.org/abs/2412.20235}
\BIBentrySTDinterwordspacing

\bibitem{nafi2024diffusionbasedapproachesmedicalimage}
\BIBentryALTinterwordspacing
A.~al~Nomaan~Nafi, M.~A. Hossain, R.~H. Rifat, M.~M.~U. Zaman, M.~M. Ahsan, and
  S.~Raman, ``Diffusion-based approaches in medical image generation and
  analysis,'' 2024. [Online]. Available: \url{https://arxiv.org/abs/2412.16860}
\BIBentrySTDinterwordspacing

\bibitem{adhikari2024parameterefficientfinetuningimprovedconvolutional}
\BIBentryALTinterwordspacing
B.~Adhikari, P.~Kulung, J.~Bohaju, L.~K. Poudel, C.~Raymond, D.~Zhang, U.~C.
  Anazodo, B.~Khanal, and M.~Shakya, ``Parameter-efficient fine-tuning for
  improved convolutional baseline for brain tumor segmentation in sub-saharan
  africa adult glioma dataset,'' 2024. [Online]. Available:
  \url{https://arxiv.org/abs/2412.14100}
\BIBentrySTDinterwordspacing

\bibitem{parida2024adultgliomasegmentationsubsaharan}
\BIBentryALTinterwordspacing
A.~Parida, D.~Capellán-Martín, Z.~Jiang, A.~Tapp, X.~Liu, S.~M. Anwar, M.~J.
  Ledesma-Carbayo, and M.~G. Linguraru, ``Adult glioma segmentation in
  sub-saharan africa using transfer learning on stratified finetuning data,''
  2024. [Online]. Available: \url{https://arxiv.org/abs/2412.04111}
\BIBentrySTDinterwordspacing

\end{thebibliography}
\bibliographystyle{ieeetr}

\end{document}